\documentstyle[11pt,paspconf,psfig]{article}

\begin{document}

\newcommand\tna{{TN~J1338$-$1942}}
\newcommand\tnb{{TN~J0924$-$2201}}
\newcommand\cf{{c.f.,~}}
\newcommand\eg{{e.g.,~}}
\newcommand\etal{{et al.~}}
\newcommand\ie{{i.e.,~}}
\newcommand\kms{\ifmmode {\rm\,km\,s^{-1}}\else
    ${\rm\,km\,s^{-1}}$\fi}
\newcommand\lya{Ly$\alpha$}
\newcommand\Lya{Ly$\alpha$}
\newcommand\CIV{\hbox{C~$\rm IV$}~$\lambda$~1549}
\newcommand\HeII{\hbox{He~$\rm II$}~$\lambda$~1640}
\newcommand\OII{[\hbox{O~$\rm II$}]~$\lambda$~3727}
\newcommand\OIII{[\hbox{O~$\rm III$}]~$\lambda$~5007}
\newcommand\arcdeg{$^{\circ}$}
\newcommand\minpoint{\ifmmode \rlap.{^{\prime}}\else
    $\rlap.{^{\prime}}$\fi}
\newcommand\secpoint{\ifmmode \rlap.{^{\prime\prime}}\else
    $\rlap.{^{\prime}}$\fi}
\def\spose#1{\hbox to 0pt{#1\hss}}
\newcommand\simlt{\mathrel{\spose{\lower 3pt\hbox{$\mathchar"218$}}
     \raise 2.0pt\hbox{$\mathchar"13C$}}}
\newcommand\simgt{\mathrel{\spose{\lower 3pt\hbox{$\mathchar"218$}}
     \raise 2.0pt\hbox{$\mathchar"13E$}}}
\newcommand\nature{{Nature}}
\newcommand\aasup{{A\&AS}}

\title{Ultra--Steep Spectrum Radio Galaxies at Hy Redshifts}

\author{Wil van Breugel, Carlos De Breuck \altaffilmark{1} \& 
Adam Stanford \altaffilmark{2}}
\affil{Inst. of Geophysics \& Planetary Physics, LLNL, Livermore,
CA 94550}

\author{Huub R\"ottgering \& George Miley} 
\affil{Leiden Observatory, Leiden, The Netherlands}

\author{Daniel Stern \altaffilmark{3}}
\affil{Astronomy Department, University of California,
    Berkeley, CA 94720}

\author{Dante Minniti}
\affil{P. Universidad Catolica, Santiago 22, Chile}

\author{Chris Carilli}
\affil{National Radio Astronomy Observatory, VLA, 
Soccoro, NM 87801}

\altaffiltext{1}{Presently at: Leiden Observatory,
Leiden, The Netherlands}
\altaffiltext{2}{University of California at Davis, CA 95616} 
\altaffiltext{3}{Presently at: JPL, 4800 Oak Grove Drive, Pasadena, CA 91109}

\begin{abstract}
Radio sources have traditionally provided convenient beacons for 
probing the early Universe. Hy Spinrad was among the first of the tenacious 
breed of observers who would attempt to obtain optical identifications 
and spectra of the faintest possible `radio galaxies' to investigate the 
formation and evolution of galaxies at hy redshift. 
Modern telescopes and instruments 
have made these tasks much simpler, although not easy, and here we 
summarize the current status of our hunts for hy redshift 
radio galaxies (HyZRGs) using radio spectral and near--IR selection. 
\end{abstract}

\keywords{radio galaxies, galaxy formation, black holes, high redshift}

\section{Hy Z Radio Galaxies: Why?}

The first optical identifications of (bright) radio sources with (faint)
galaxies were made when Hy Spinrad was still a teenager (Windhorst 1999;
these proceedings). After that it was soon realized that the `invisible'
universe of radio sources provided convenient beacons to locate very
distant galaxies and thus might be used to study their formation and
evolution. As so eloquently described by several of Hy's colleagues,
collaborators and ex--students in these proceedings, he became an early
key player in these distant galaxy hunts.

For most radio astronomers in those days the Universe stopped at the
POSS limits. Surely, many radio sources could not be identified, but so
what? It just confirmed that the Universe was bigger than the biggest
optical telescope, but ... not bigger than the biggest {\it radio}
telescope! Occasionally Hy would write letters to Leiden Observatory
radio astronomers with requests of radio maps and accurate positions.
When provided he would spend many hours using one of the world's then
finest telescopes at Lick Observatory, to obtain optical
identifications and spectra of these HyZRG candidates (presumably
squeezed in between observations of standard stars for Jim Liebert;
these proceedings). How foolish this seemed, to some of us (WvB).

Since then extraordinary progress in the development of optical and
near--IR detectors, larger telescopes, and better selection techniques
have resulted in discoveries of radio galaxies at increasingly hyer
redshifts. Paradoxically this task was eased by the discovery, first by
Hy and his collaborators, that the \lya\ emission line was very strong
in radio galaxies and could easily be detected in $< 1$ hr integrations
with 3m--class telescopes (Spinrad \etal 1985), provided that the
redshifts would be hy enough ($z > 1.6$) so that \lya\ would enter the
observable optical window.

It has now become clear that HyZRGs are both a boon and a curse for
students of galaxy evolution.  A boon because the near--IR Hubble $K-z$
relation for radio sources appears well represented by the predicted
`passive' evolution of massive ($5 - 10$ L$_\star$) galaxies with hy
formation redshifts (Lilly \& Longair 1984; Eales \etal 1997; Best \etal
1998; van Breugel \etal 1999), despite the effects of k--correction
and morphological evolution (van Breugel \etal 1998). No matter the
reason for this relationship, it suggests that radio sources might be
used to find massive galaxies and their likely progenitors out to very
hy redshift. This method was first successfully used by Lilly (1988)
to identify the HyZRG B2~0902$+$34 at $z = 3.395$.  Deep spectroscopic
observations of a few relatively weak radio sources (but still of the
powerful double FRII class !), where the AGN do not affect the rest--frame
UV as much (see below), have shown directly examples of radio galaxies
at $z \sim 1.5$ with old ($\simgt 3.5 - 4.5$ Gyr) stellar populations
with implied formation redshifts $z_f > 10$ (\eg Spinrad \etal 1997).

For galaxy formation studies HyZRGs are also cursed because their
structures are aligned with their associated radio sources, suggesting
that the collimated outflow and ionizing radiation from their AGN
profoundly affect their parent galaxy host appearances at UV, blue and
green wavelengths (McCarthy 1999, and Dey 1999; these proceedings). 
HyZRGs are very well suited for studying the effects of
powerful AGN on ambient dense gas, including induced star formation
(\eg Bicknell \etal 1999), and may even be used as searchlights to
investigate the properties of proto--galactic material in the early
Universe (Cimatti \etal 1997; Villar--Mart\'\i n \etal 1997).

Recent cosmological theories are providing additional incentives to use
radio galaxies as probes to study the early Universe.  Within standard
Cold Dark Matter scenarios the formation of galaxies is a hierarchical and
biased process. Large galaxies are thought to grow through the merging of
smaller systems, and the most massive objects form in over--dense regions,
which will eventually evolve into the clusters of galaxies seen today
(\eg\ White 1997).  It has also been suggested that the first massive
black holes may grow in similar hierarchical fashion together with
their parent galaxies (\eg\ Kauffmann and Haehnelt 1999) or, because of
time scale constraints, may precede galaxy formation and be primordial
(\eg\ Loeb 1993). To confront these theories it is therefore of great
interest to find the progenitors of the most massive galaxies and their
AGN (active massive black holes) at the hyest possible redshifts and
to study their properties and cosmological evolution.

While optical, `color--dropout' techniques have been successfully used
to find large numbers of 'normal' young galaxies (without dominant AGN)
at redshifts even surpassing those of quasars and radio galaxies (Weymann
\etal 1998; Hu \etal 1999), the radio and near--infrared
selection technique has the additional advantage that it is unbiased
with respect to the amount of dust extinction. HyZRGs are therefore
also important laboratories for studying the large amounts of dust (\eg\
Ivison \etal 1998) and molecular gas (Papadopoulos \etal 1999), which are
observed to accompany the formation of the first forming massive galaxies.
Indeed, a significant part of the scientific rationale for building
future large mm-arrays is based on the expectation that to understand
galaxy formation will ultimately require understanding of their cold
gas and dusty environments.

\begin{figure}
\centerline{\psfig{file=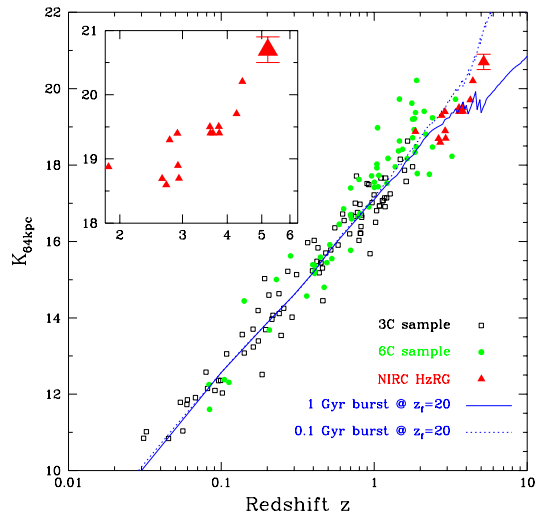,width=8cm}}
\caption{The Hubble $K-z$ diagram for HyZRGs. Filled triangles are Keck
measurements of HzRGs from van Breugel \etal (1998), the large triangle
is \tnb\ at $z = 5.19$, 
and all other photometry is from Eales \etal (1997).  Two
stellar evolution models from Bruzual \& Charlot (1999), normalized at
$z<0.1$, are plotted, assuming parameters as shown.}
\end{figure}

Finally, it has been claimed that the (co--moving) space densities of
the most powerful radio galaxies and quasars were much hyer near $z
\sim 2$, but that they drop off precipitously at even hyer redshifts
(Dunlop and Peacock 1990; Shaver \etal 1996). However, using recently
completed studies of moderately faint radio galaxies Jarvis \etal
(1999; these proceedings) have argued that here is {\it no} such
evidence for a redshift cut--off and that these previous results have
been biased due to unknown radio K--correction (radio spectral
index) trends and associated selection effects. 

For all these reasons HyZRGs have become very useful probes of the early
Universe. Unfortunately in complete, flux--limited samples the vast
majority of the sources will be relatively nearby, or at only modest
hy redshifts. However, by employing the well-known `red radio color'
selection technique and choosing sources with ultra--steep radio spectra
one can bypass most of the `local' radio source population and efficiently
identify radio galaxies at extremely hy redshift.

\section{The Ultimate Ultra--Steep Spectrum Source Sample}

\begin{table}[t]
\caption{Radio Surveys}
\footnotesize
\begin{tabular}{lccc}
\hline
\hline
\\
 & WENSS & TEXAS & MRC \\
\\
\hline
\hline
\\
Frequency (MHz) & 325 & 365 & 408 \\
Sky region & $\delta >$ +29\arcdeg & $-$35\fdg7 $< \delta <$ +71\fdg5 & $-$85\arcdeg $< \delta <$ +18\fdg5 \\
\# of sources & 229,576 & 67,551 & 12,141 \\
Resolution & $ 54\arcsec \times$ 54\arcsec cosec$\delta$ & 10\arcsec & $2\farcm62 \times 2\farcm86 \sec (\delta - 35\fdg5)$ \\
Position uncertainty & 1\farcs5 & 0\farcs5---1\arcsec & 8\arcsec \\
(strong sources) & & & \\
RMS noise & $\sim$4 mJy & 20 mJy & 70 mJy \\
Flux density limit & 18 mJy & 150 mJy & 670 mJy \\
\\
\hline
\hline
\\
 & NVSS & FIRST\tablenotemark{a} & PMN \\
\\
\hline
\hline
\\
Frequency (MHz) & 1400 & 1400 & 4850 \\
Sky region & $\delta > -$40\arcdeg & $b>45$ & $-$87\fdg5 $< \delta <$ +10\arcdeg \\
\# of sources & 1,814,748 & 550,000 & 50,814 \\
Resolution & 45\arcsec $\times$ 45\arcsec & 5\arcsec $\times$ 5\arcsec & 4\minpoint2 \\
Position uncertainty & 1\arcsec & 0\farcs1 & $\sim$45\arcsec \\
(strong sources) & & & \\
RMS noise & 0.5 mJy & 0.15 mJy & $\sim$8 mJy \\
Flux density limit & 2.5 mJy & 1 mJy & 20 mJy \\
\end{tabular}
\tablenotetext{a}{Still in progress.}
\end{table}

With the advent of several new, large radio surveys (Table 1) such as
the 325 MHz WENSS (Rengelink \etal\ 1997), the 365 MHz TEXAS (Douglas
\etal\ 1996), the 1.4 GHz NVSS (Condon \etal\ 1998), the 1.4 GHz FIRST
(Becker \etal\ 1995), MRC 408~MHz (Large \etal 1981) and the 4.85~GHz
PMN (Griffith \etal 1993) surveys it is now possible for the first
time to define a large sample of USS sources with extremely steep
radio continuum spectra ($\alpha \le -1.3$, Fig. 2), and using 10 --
100 times lower flux density limits than has been possible before
(Chambers \etal 1996; R\"ottgering \etal\ 1994; Blundell \etal 1998).

Using these surveys we constructed 3 sub--samples, covering different
regions of the radio sky using the deepest low and high frequency surveys
available in  each area (Table 2). Our largest, and most complete sample
is based on the WENSS survey at $\delta > 29$\arcdeg, together with the
NVSS and FIRST surveys. In the remaining area covered by the northern
hemisphere radio telescopes, we used the Texas survey at low frequencies,
which produces a similar sample but at a higher flux density and is
less complete.  We also used two southern surveys to construct the first
USS sample in the deep southern sky. More details about the samples are
given in De Breuck \etal\ (2000$a$).

\begin{figure}
\begin{minipage}{6.56cm}
\centerline{\psfig{file=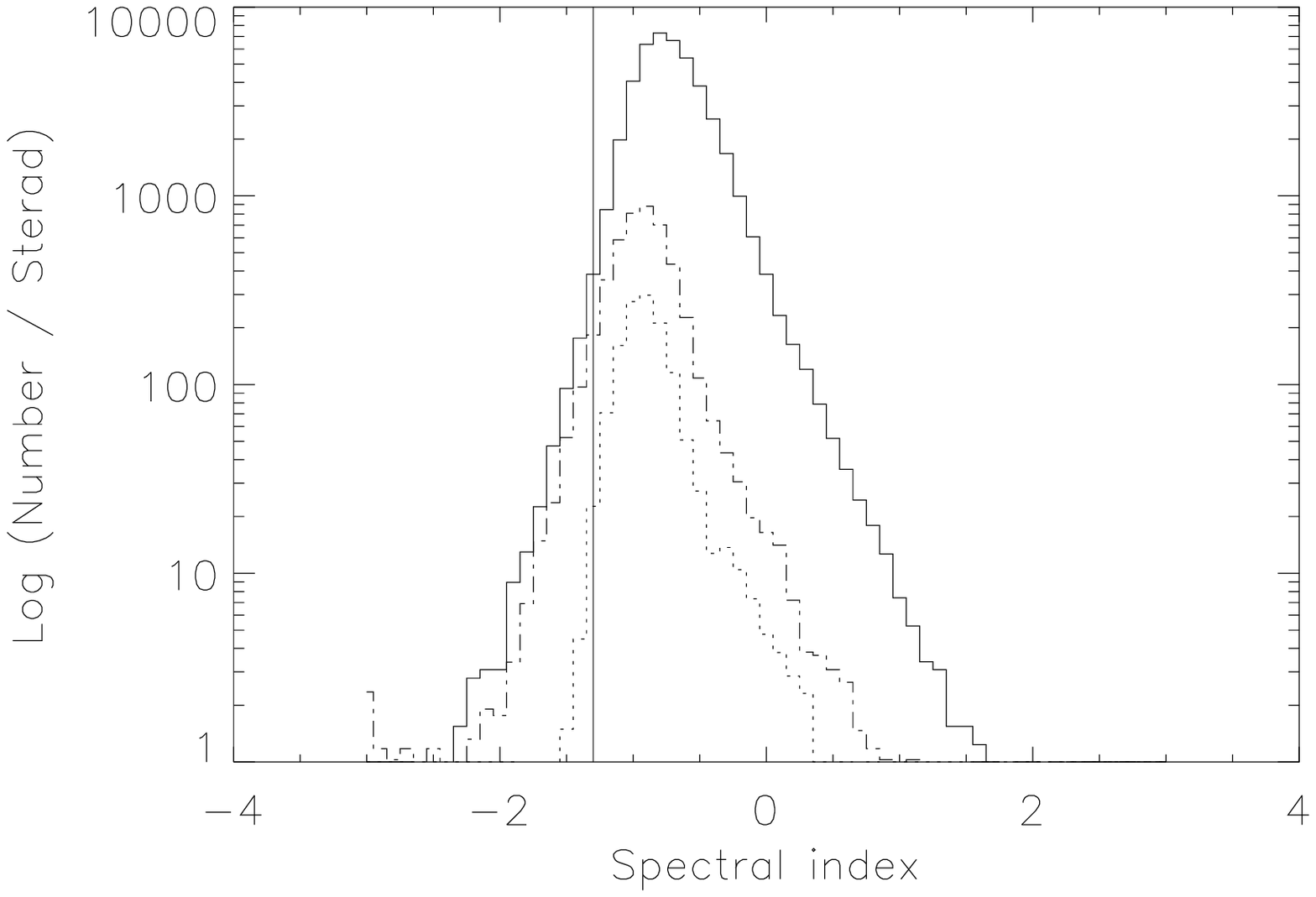,width=8cm}}
\caption{Logarithmic spectral index distribution for WENSS--NVSS (full line), Texas--NVSS (dot--dash line) and MRC--PMN (dotted line). The vertical line indicates the --1.3 cutoff used in our spectral index selection.}
\end{minipage}
\begin{minipage}{6.66cm}
\centerline{\psfig{file=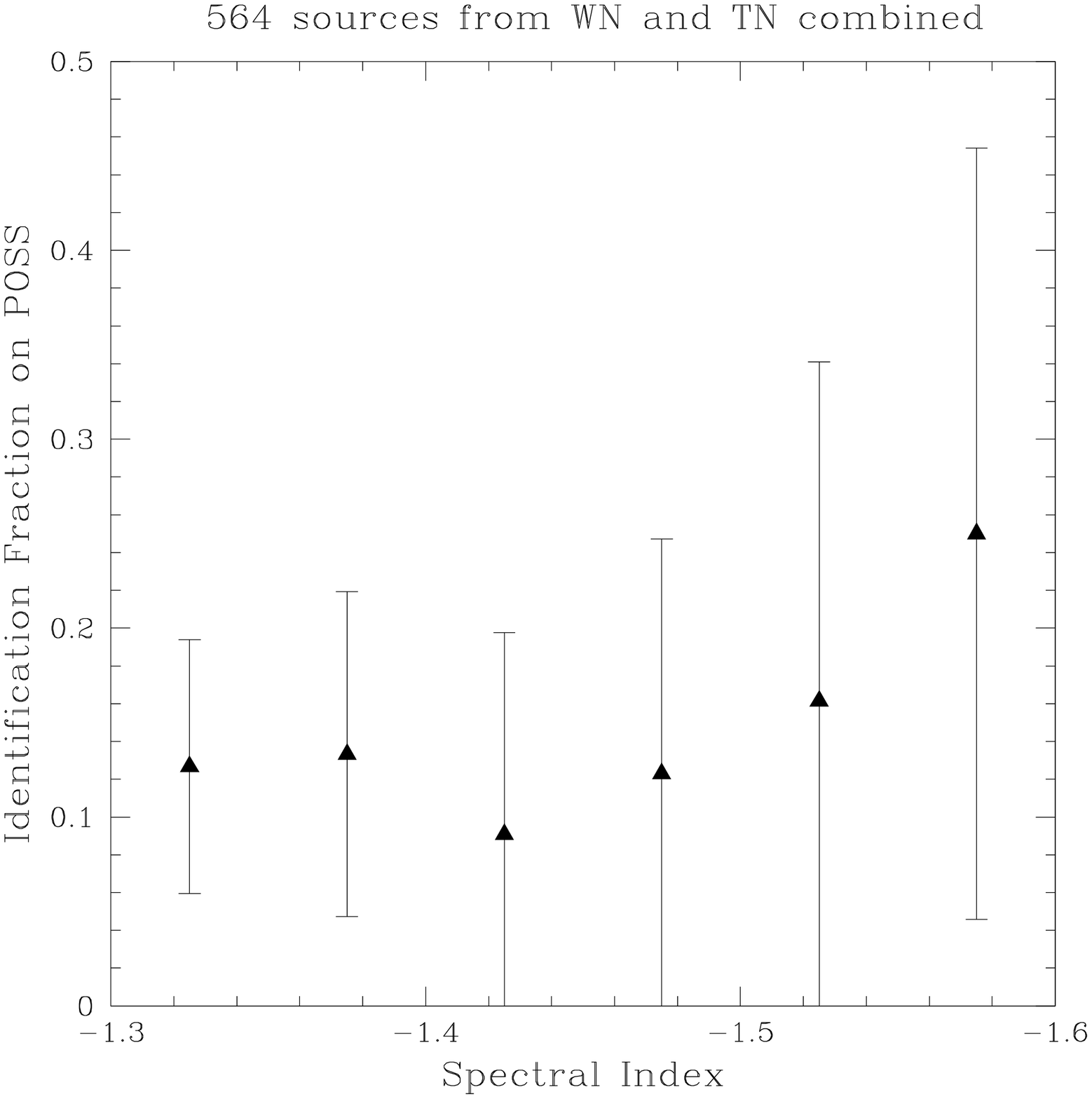,width=6cm}}
\caption{Identification fraction on the POSS as a function 
of spectral index for the combined WN and TN sample.} 
\end{minipage}
\end{figure}

During the course of our optical and near--IR imaging and optical
spectroscopy programs we have fine--tuned our selection technique.
Previously it had been found that the identification fraction of radio
galaxies decreases with spectral index (Tielens \etal 1979; R\"ottgering
\etal 1995), which provided the rationale for using the USS source
selection technique.  With our `hyper-steep' radio spectrum selection
($\alpha \le -1.3$) most sources remain unidentified, at least on the
POSS ($R \simlt 20$). Only $\sim$ 15\% of the sources can be identified,
usually with bright galaxy clusters, as indicated by the frequent
overdensity of galaxies around them, and by X-ray detections (De Breuck
\etal 2000$a$). This identification fraction appears to be independent of
spectral index (Fig. 3), in support of the idea that these are mostly
foreground objects.

This also explained why our initial optical imaging campaign on
3m--4m--class telescopes ($R \simlt 24$) was not very succesful in
finding $R-$band identifications. Furthermore, for the typically
expected $R - K \sim 4$ values of HyZRGs, it would even be a challenge
to detect most HyZRGs in the near--IR at Lick Observatory.  We
therefore decided to entirely skip the optical identification program
and go straight to near--IR imaging at the Keck I telescope. This 
has produced, to date, a 100\% identification rate 
with good photometric magnitudes to select HyZRG candidates using the 
Hubble $K-z$ diagram (Fig. 1).

We have now spectroscopically observed 30 faint USS HyZRG candidates
with the following results.  Only 5 of the sources have $z<2$, 7 have
$2<z<3$, 7 have $3<z<4$ and 2 sources have $z>4$, including one at $z >
5$. At least 3 sources failed to yield redshifts, and were not detected
in the continuum, despite $\sim$ 1 hr integrations with LRIS, and may
be at record hy redshifts, or are extremely obscured.  We also found 6
sources with only a continuum detection and no emission--lines. These
were all extremely compact USS sources, and may be moderately hy redshift
($1 < z < 3$) BL Lac objects, `emission--line free quasars' (\cf Fan
\etal 1999), or even pulsars (which typically have $\alpha_{radio} \sim -1.6$,
Kaplan \etal 1998, and are faint optically; Martin \etal 1998).

\begin{table}
\caption{USS samples}
\small
\begin{tabular}{lcccc}
\hline
\hline
Sample & Density & Spectral Index & Flux Limit & \# of Sources \\
 & sr$^{-1}$ & & mJy & \\
\hline
\hline
WN & 151 & $\alpha_{325}^{1400} \le -1.30$ & $S_{1400} >$ 10 & 343 \\
TN & 48\tablenotemark{a} & $\alpha_{365}^{1400} \le -1.30$ & $S_{1400} >$ 10 & 268 \\
MP & 26 & $\alpha_{408}^{4800} \le -1.20$ & S$_{408} > 700$; S$_{4850} > 35$ & 58 \\
\end{tabular}
\tablenotetext{a}{Due to the characteristics of the Texas survey, the TN sample is only $\sim 30\%$ complete.}
\end{table}

\section{The Highest-Redshift Radio Galaxies}

\subsection{TN~J1338$-$1942 at $z = 4.11$}

\begin{figure}
\begin{minipage}{6.6cm}
\centerline{\psfig{file=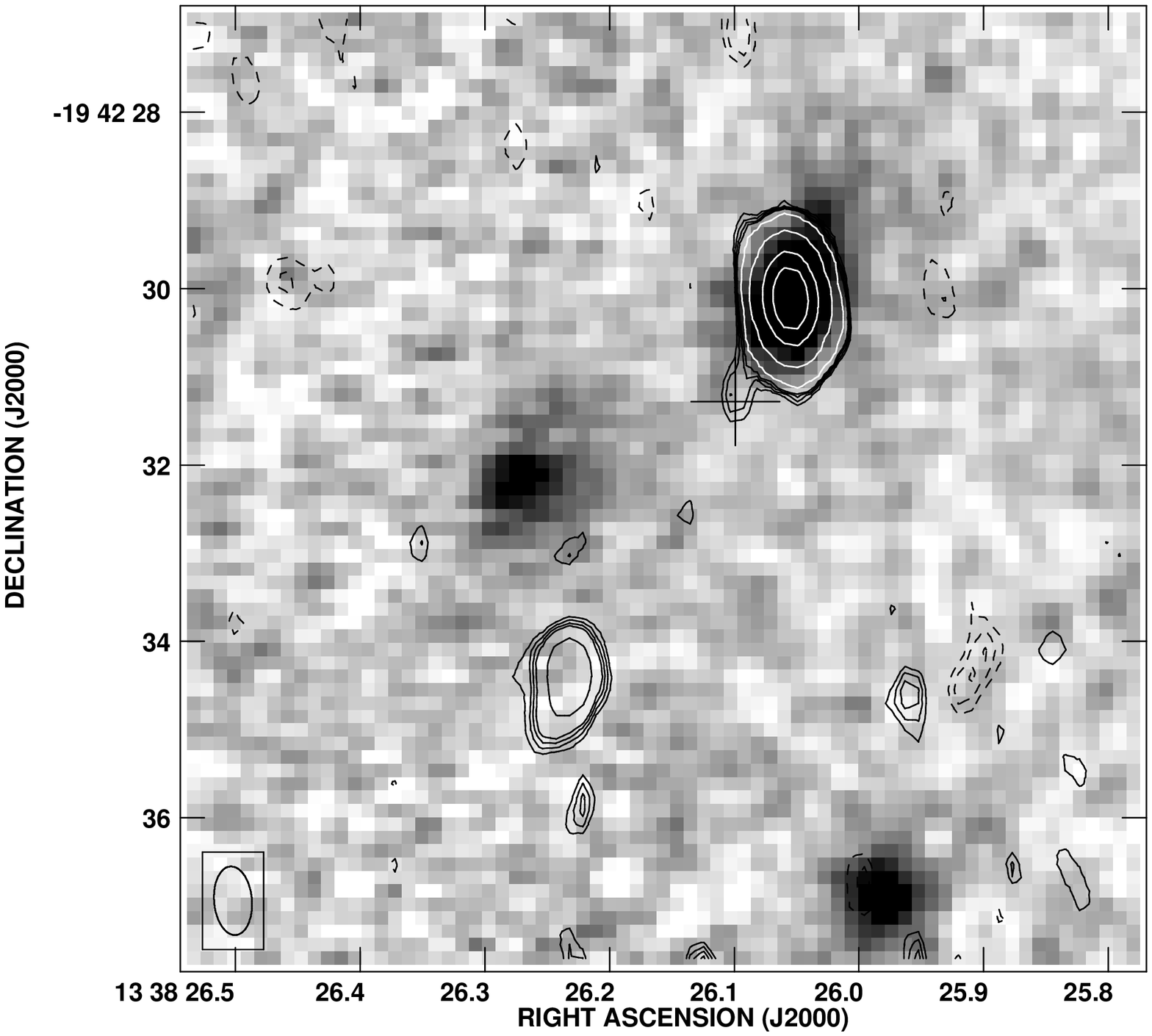,width=6cm,angle=-90}}
\caption{4.85~GHz VLA radio contours overlaid on a Keck $K-$band
image of \tna.}
\end{minipage}
\begin{minipage}{6.6cm}
\centerline{\psfig{file=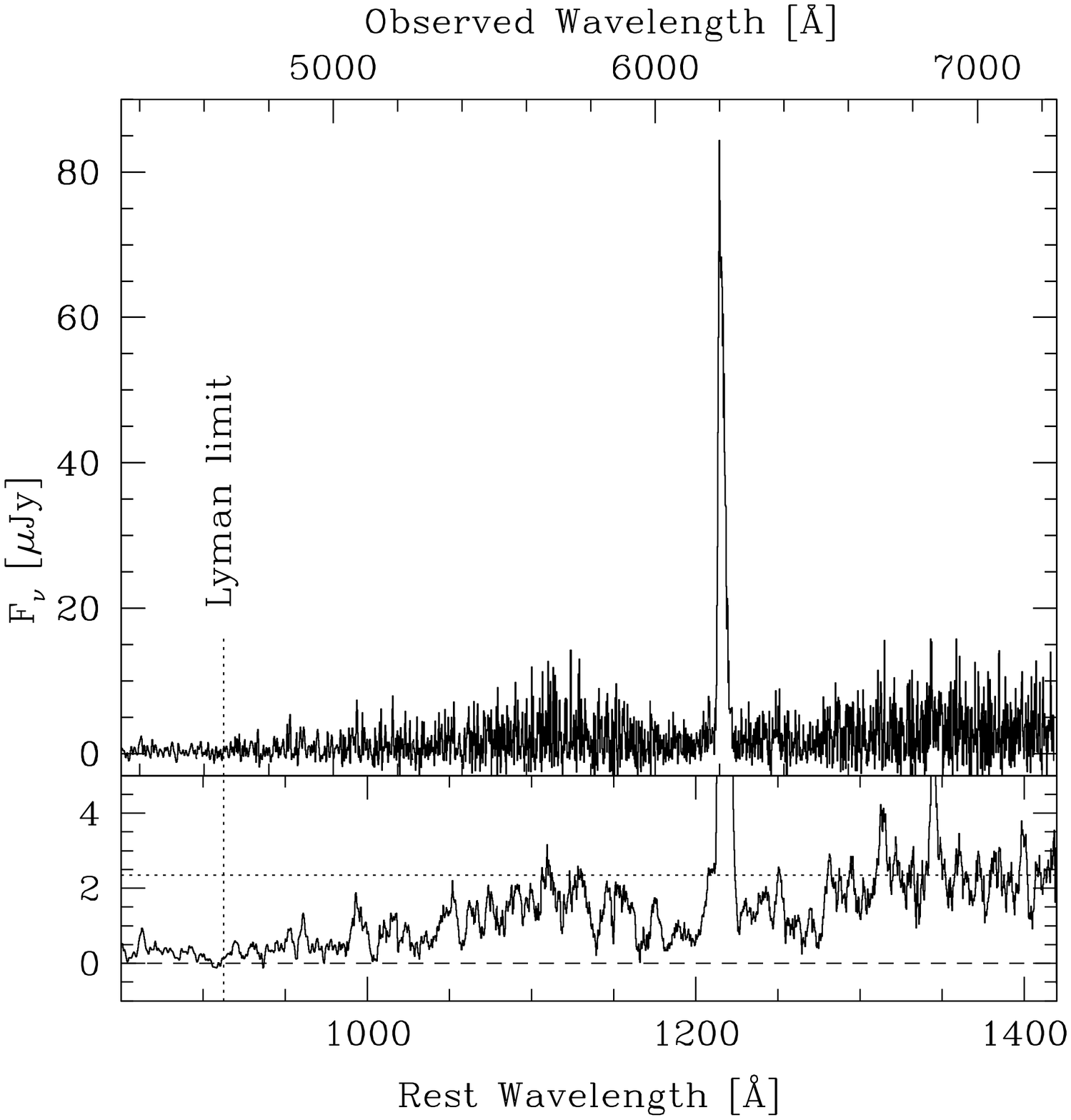,width=6cm}}
\caption{VLT spectrum of \tna at $z = 4.11$.} 
\end{minipage}
\end{figure}

The first $z > 4$ USS radio galaxy discovered by us was
TN~J1338$-$1942. The initial identification was made with the ESO
3.6m at R--band, and subsequent spectroscopy with that same telescope
showed that the radio galaxy has a redshift of $z=4.11 \pm 0.02$, based
on a strong detection of \lya, and weak confirming \CIV\ and \HeII\
(De Breuck \etal 1999$a$).

Subsequently we obtained a deep K--band image (rest--frame B--band) at
Keck, shown in Figure 4 overlaid with a VLA radio image (De Breuck
\etal 1999$b$).  The radio source is an asymmetric double, with a very
bright NW hotspot ($S_{4.7GHz}$ = 22 mJy; $\alpha^{\rm 8.5~GHz}_{\rm
4.7~GHz} \sim -1.6$) coincident with the peak of the $K-$band
emission. This hotspot has a very faint radio companion with a flatter
spectrum ($S_{4.7GHz}$ = 0.3 mJy; $\alpha^{\rm 8.5~GHz}_{\rm 4.7~GHz}
\sim -1.0$) at 1\farcs4 to the SE which we identify as the probable
nucleus.  Thus the AGN and rest--frame optical (continuum and \lya)
emission may not be co--centered.  This resembles the $z = 3.800$
radio galaxy 4C41.17 (van Breugel \etal 1999). A possible reason for
this might be that the central region is obscured by dust. Such
asymmetric radio sources are not uncommon, even in the local Universe,
and are usuallly thought to be due to strong interaction of one of its
radio lobes with very dense gas (\eg McCarthy \etal 1991; Feinstein
\etal 1999).

A high signal--to--noise spectrum was also obtained with the VLT Antu
telescope (De Breuck \etal 1999$b$). The spectrum is dominated by the
bright \Lya\ line ($W_{\rm Ly\alpha}^{\rm rest}$ = 210\AA) which shows
deep and broad ($\sim 1400$ \kms) blue--ward absorption.  The latter is
probably due to resonant scattering by cold HI gas in a turbulent halo
surrounding the radio galaxy and has also been seen several other HyZRGs
(van Ojik \etal 1996; Dey 1999).  The continuum is relatively
bright ($F_{1400} \sim 2 \mu$Jy) and if all due to young O--B stars
this would imply a total SFR of several hundred M$_{\odot}$/yr,
resembling 4C41.17, and suggesting that \tna\ may be another HyZRG in
which induced star formation might occur (\cf Bicknell \etal 1999).

In Table 3 we have listed the \lya\ properties of the known 7 most
distant radio galaxies for which high quality optical slit
spectroscopy data taken with Keck or the VLT are available.  We have
assumed $H_0= 65$~km~s$^{-1}$Mpc$^{-1}$, $q_0$=0.15, and
$\Lambda=0$. The \Lya\ fluxes are as measured \ie uncorrected for
blue--ward absorption.  \tna\ is the most luminous \lya\ galaxy and,
after 4C41.17, also the brightest (in similar apertures).  In all
cases the brightest \lya\ emission occurs on scale sizes of
$1^{\prime\prime} - 2^{\prime\prime}$, comparable to those of the
brightest radio structures. 4C41.17 is known to have a very extended
halo (Chambers \etal 1990) and the total size quoted is a lower limit,
based on the deep (9 hrs) Keck spectropolarimetry data from Dey \etal
(1997).

\subsection{TN~J0924$-$2201 at $z = 5.19$}

TN~J0924$-$2201 is one of the steepest spectrum sources in our USS sample
($\alpha_{\rm 365 MHz}^{\rm 1.4 GHz} = -1.63$) and therefore was one of
our primary targets for near--IR identification.  A deep K--band image at
Keck showed indeed a very faint ($K = 21.3 \pm 0.3$), multi--component
object at the position of the small ($1\farcs2$) radio source (Fig. 6).
The expected redshift on the basis of the $K-z$ diagram was $z > 5$,
and spectroscopic observations at Keck showed that this was indeed
the case, based on a single emission line at $\lambda \sim 7530$ \AA\
which we identified as \lya\ at $z = 5.19$ (van Breugel \etal 1999;
none of the $z > 5$ galaxies have more than one line detection).

\begin{figure}
\begin{minipage}{6.2cm}
\vspace{-1cm}
\vspace{-3cm}
\centerline{\psfig{file=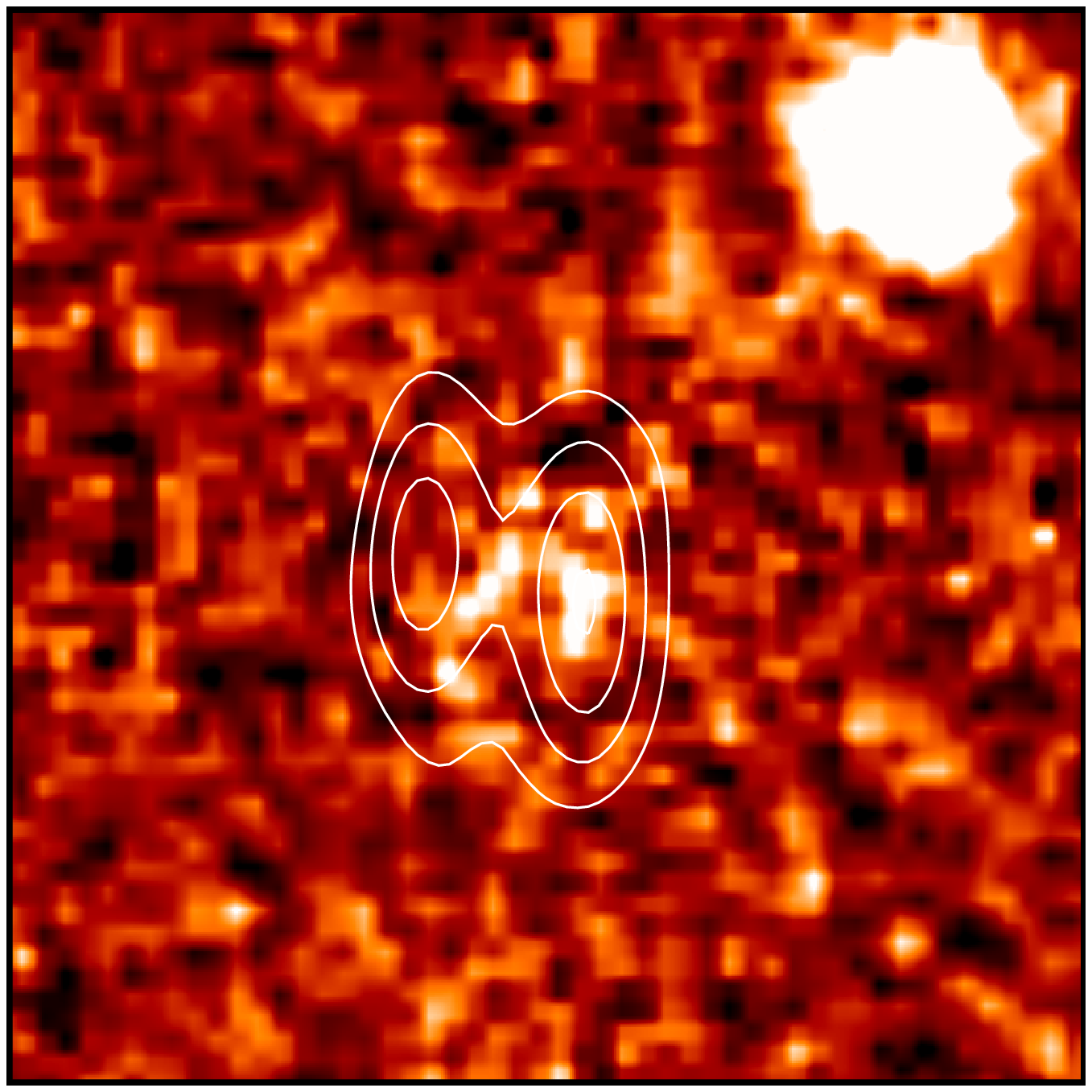,width=8cm}}
\vspace{-1cm}
\caption{Keck/NIRC $K$-band image of \tnb, with radio contours superposed.
}
\end{minipage}
\begin{minipage}{7cm}
\vspace{-1cm}
\centerline{\psfig{file=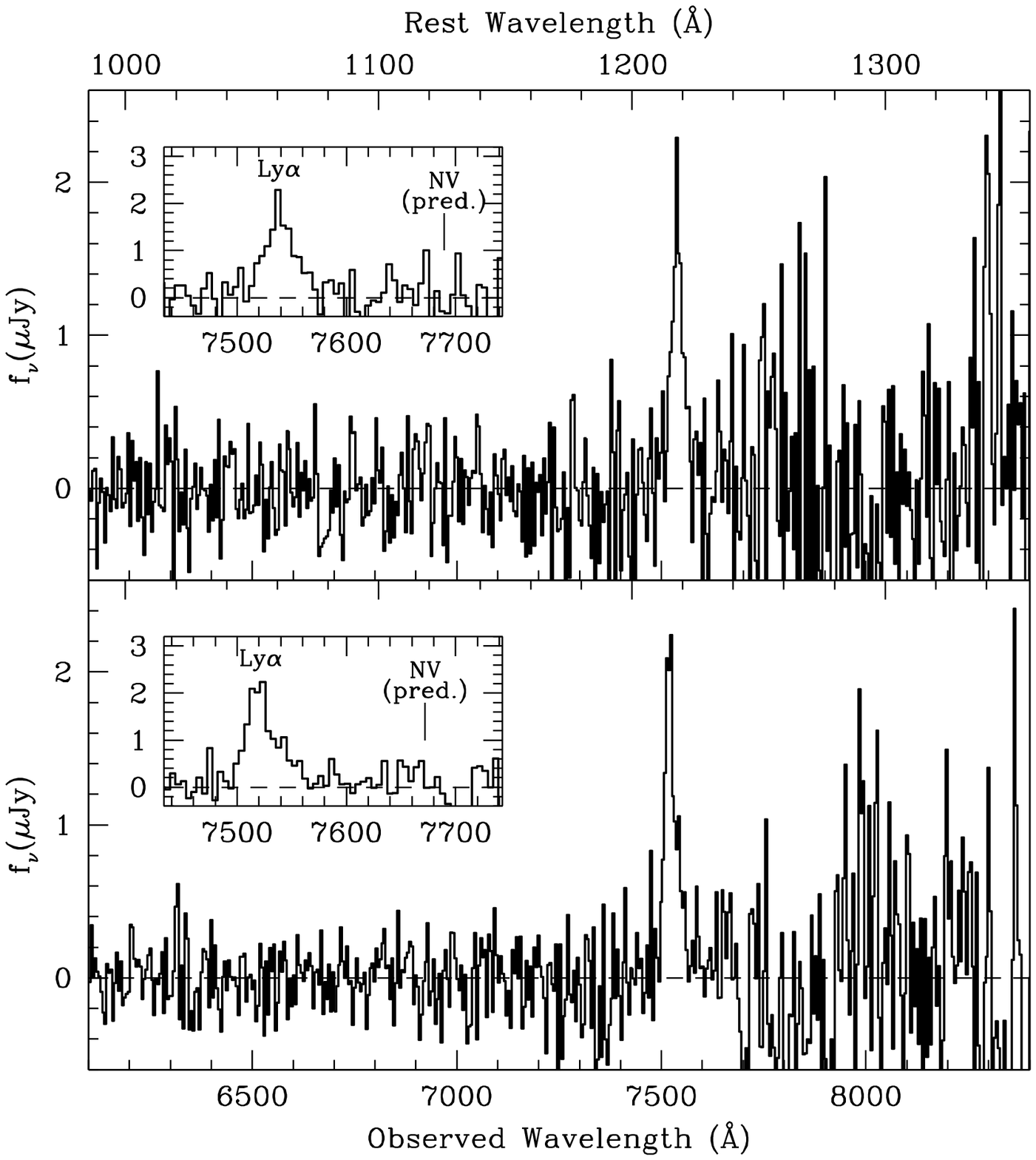,width=7cm}}
\caption{Keck spectra of \tnb at $z = 5.19$ on two different nights.}
\end{minipage}
\end{figure}

Among all radio selected HyZRGs \tnb\ is fairly typical in radio
luminosity, equivalent width and velocity width (Table~2). It does have
the steepest radio spectrum, consistent with the $\alpha - z$ relationship
for powerful radio galaxies (\eg\ R\"ottgering \etal 1997), and also has
the smallest linear size.  The latter may be evidence of its `inevitable
youthfulness' or a dense confining environment, neither of which would be
surprising because of its extreme redshift (Blundell \& Rawlings 1999;
van Ojik \etal 1997).  Among the radio selected HyZRGs \tnb\ appears
underluminous in \lya, together with 8C~1435$+$63, which might be caused
by absorption in an exceptionally dense cold and dusty medium. Evidence for
cold gas and dust in several of the most distant HyZRGs has been found
from sub--mm continuum and CO--line observations (\eg Ivison \etal 1998;
Papadopoulos \etal 1999).

The second hyest redshift radio galaxy currently known listed in
Table 3 is VLA~J123642+621331 at $z = 4.42$ (Waddington \etal
1999). This source was not USS selected and therefore provides an
interesting alternate view on the possible selection effects of our
method of finding HyZRGs. The source is an asymmetric double and
although its radio luminosity is about a factor 1000 times lower than
that of its much more luminous brothers at similar redshifts, it still
qualifies as a FRII--type, though with a radio luminosity close to the
FRI / FRII break at 408 MHz ($P_{408} \sim 3.2 \times 10^{26}$W
Hz$^{-1}$). Its radio spectrum is steep ($\alpha_{8.4GHz}^{1.4GHz}
\sim -1.0$, using the flux densities given by Waddington \etal), but
not as steep as our USS selected HyZRGs, and the \lya\ luminosity is a
factor 5 -- 10 times less. Apart from the luminosity these properties
are not hugely different from expected on the basis of radio selection
and indicate that less extreme steep spectrum selected samples
($\alpha < -1.0$) at much lower flux densities ($\simlt 1 mJy$) might
be used to find many more (hundreds, thousands ?)  HyZRGs at very hy 
redshifts (though with much lower efficency, we suspect, than USS
selected samples).

\begin{table}
\caption{Physical Parameters of the Highest HyZRGs} \label{tbl-1}
\begin{center}
\small
\begin{tabular}{crrrrrrrc}
\hline
\hline
Name & $z$ & $L_{\rm Ly\alpha}$\tablenotemark{a} & $L_{\rm 365}$\tablenotemark{a}
& $\alpha_{\rm 365}^{\rm 1400}$ & $W_{\rm Ly\alpha}^{\rm
rest}$ & $\Delta_{\rm Ly\alpha}$\tablenotemark{a} &
\multicolumn{1}{c}{Size\tablenotemark{a}} & Ref.\tablenotemark{b} \\
\hline
\tableline
TN~J0924$-$2201 &  5.19 & 1.3 & 7.5 & $-$1.63 &  $>$115 & 1500 &  8 & WvB99 \\
VLA~J1236+6213 & 4.42 & 0.2 & 0.0035 & $-$0.96 &  $>$50 &  440 &    & Wad99 \\
6C~0140$+$326   &  4.41 &  16 & 1.3 & $-$1.15 &     700 & 1400 & 19 & DeB00 \\
8C~1435$+$63    &  4.25 & 3.2 & 11  & $-$1.31 &     670: & 1800 & 28 & Spin95 \\
TN~J1338$-$1942 &  4.11 &  25 & 2.3 & $-$1.31 &     200 & 1000 & 37 & DeB99 \\
4C~41.17        & 3.798 &  12 & 3.3 & $-$1.25 &     100 & 1400 & 99 & Dey97 \\
4C~60.07        &  3.79 &  16 & 4.1 & $-$1.48 &     150 & 2900: & 65 & R\"ot97 \\
\end{tabular}
\end{center}
\tablenotetext{a}{In units of $10^{43}$ erg s$^{-1}$ ($L_{\rm Ly\alpha}$),
$10^{36}$ erg s$^{-1}$ {Hz}$^{-1}$ ($L_{\rm 365}$),
restframe km $s^{-1}$, kpc respectively)}
\tablenotetext{b}{Most recent references quoted only:
WvB99 = van Breugel \etal\ (1999); 
Wad99 = Waddington \etal\ (1999);
DeB00 = De Breuck \etal\ (2000$b$);
Spin95 = Spinrad \etal\ 1995;
DeB99 = De Breuck \etal\ (1999$a,b$);
Dey97 = Dey \etal\ (1997); 
R\"ot97 = R\"ottgering \etal (1997).}
\end{table}

Our observations of \tnb\ extend the Hubble $K-z$ diagram for powerful
radio galaxies to $z = 5.19$, as shown in Figure~1.  Simple stellar
evolution models are shown for comparison.  Despite the enormous
$k$--correction effect (from $U_{\rm rest}$ at $z = 5.19$ to
$K_{\rm rest}$ at $z = 0$) and strong morphological evolution (from
radio--aligned to elliptical structures), the $K-z$ diagram remains a
powerful phenomenological tool for finding radio galaxies at extremely
hy redshifts. Deviations from the $K-z$ relationship may exist (Eales
\etal 1997; but see McCarthy 1999), and scatter in the $K-z$ values
appears to increase with redshift, but this may in part be caused by
limited signal-to-noise or emission--line contamination.

The clumpy $U_{\rm rest}$ morphology resembles that of other HyZRGs (van
Breugel \etal 1998; Pentericci \etal 1998) and if it is dominated by
star light we derive a SFR of $\sim$200 M$_\odot$ yr$^{-1}$, without any
correction for extinction, which may be a factor of several.  \tnb\ may
be a massive, active galaxy in its formative stage, in which the SFR is
boosted by jet--induced star formation (Dey \etal 1997; van Breugel \etal
1999; Bicknell \etal 1999).  For comparison other, `normal' star forming
galaxies at $z > 5$ have 10 -- 30 times lower SFR ($\sim 6 - 20 M_{\odot}
yr^{-1}$; Dey \etal 1998; Weymann \etal 1998; Spinrad \etal 1998).

At $z = 5.19$ \tnb\ is currently the most distant AGN known, surpassing
even quasars for the first time since their discovery 36 years ago. The
presence of AGN at such early epochs in the Universe ($< $1 Gyr in most
cosmogonies) poses interesting challenges to common theoretical wisdom,
which assumes that they are massive (billion solar mass), active black
holes. The question how these can form so shortly after the putative
Big Bang may prove even more challenging then that of the formation
of galaxies (\eg Loeb 1993; Silk \& Rees 1998).

\acknowledgments

WvB is grateful for the many interesting conversations, advice,
and wonderful collaborations he has had with Hy Spinrad and his
students over the past 15 years.  The work by W.v.B., C.D.B. and
S.A.S.\ at IGPP/LLNL was performed under the auspices of the US
Department of Energy under contract W-7405-ENG-48.  W.v.B.\ also
acknowledges support from NASA grants GO 6608, and D.S. from IGPP/LLNL
grant 98--AP017.


\begin{references}
\reference Becker, R. H., White, R. L., \& Helfand, D. J. 1995, \apj, 450, 559
\reference Bicknell, G., Sutherland, R., van Breugel, W., Dopita, M., Dey, A., Miley, G. 1999, \apj, in press, astro-ph/9909218
\reference Blundell, K. M., \etal 1998, \mnras, 295, 265
\reference Blundell, K. M., \& Rawlings, S. 1999, \nature, 399, 330
\reference Bruzual, A. G., \& Charlot, S.\ 1999, personal communication
\reference Chambers, K. C., Miley, G. K., \& van Breugel, W. J. M. 1990, \apj, 363, 21
\reference Chambers, K. C., Miley, G. K., van Breugel, W. J. M., \& Huang, J.-S. 1996, \apjs, 106, 215
\reference Cimatti, A., Dey, A., van Breugel, W., Hurt, T. \& Antonucci, R. 1997, \apj, 476, 677 
\reference Condon, J., \etal 1998, \aj, 115, 1693
\reference De Breuck, C., van Breugel, W., R\"ottgering, H., Miley, G. \& Carilli, C. 1999$a$, in 'Looking Deep in the Southern Sky' Edited by R. Morganti \& W. Couch (Springer), p.\ 246 
\reference De Breuck, C., van Breugel, W., Minniti, D., Miley, G., R\"ottgering, H., Stanford, S., \& Carilli, C. 1999$b$, \aap, in press (astro-ph/9909178) 
\reference De Breuck, C., van Breugel, W., R\"ottgering, H., \& Miley, G. 2000$a$, \aasup, submitted
\reference De Breuck, C., van Breugel, W., R\"ottgering, H., Miley, G., \& Stern, D. 2000$b$, in preparation
\reference Dey, A., van Breugel, W., Vacca, W. D., \& Antonucci, R. 1997, \apj, 449, 698
\reference Dey, A., Spinrad, H., Stern, D., Graham, J., \& Chaffee, F. 1998, \apjl, 498, L93 
\reference Dey, A. 1999, in 'The Most Distant Radio Galaxies', ed. H. R\"ottgering, P. Best \& M. Lehnert (Amsterdam: KNAW), p. 19
\reference Douglas, J. N., \etal 1996, \aj, 111, 1945
\reference Dunlop, J., \& Peacock, J. 1990, \mnras, 247, 19
\reference Eales, S., \etal  1997, \mnras, 291, 593
\reference Fan, X. \etal 1999, apjl, in press, astro-ph/9910001
\reference Feinstein, C., \etal 1999, astro-ph/9906059
\reference Griffith, M. \& Wright, A. E. 1993, \aj, 105, 1666
\reference Haehnelt, M., Natarajan, P. \& Rees. M. J. 1998, \mnras, 300, 817
\reference Hu, E., McMahon, R., \& Cowie, L. 1999, \apj, 522, 9
\reference Ivison, R. J. \etal 1998, \apj, 494, 211
\reference Kaplan, D. L., \etal 1998, \apjs, 119, 75 
\reference Kauffmann, G., \& Haehnelt, M. 1999, \mnras, in press, astro-ph/9906493
\reference Large, M. I., Mills, B. Y., Little, A. G., Crawford, D. F., \& Sutton, J. M. 1981, \mnras, 194, 693
\reference Lilly, S. J. \& Longair, M., 1984, \mnras, 211, 833
\reference Lilly, S. 1988, \apj, 333, 161
\reference Loeb, A. 1993, \apj, 403, 542
\reference Martin, C., Halpern, J., \& Schiminovich, D. 1998 \apj, 494, 211
\reference McCarthy, P., van Breugel, W., \& Kapahi, V. K. 1991, \apj, 371, 478
\reference McCarthy, P. 1999, in 'The Most Distant Radio Galaxies', ed. H. R\"ottgering, P. Best \& M. Lehnert (Amsterdam: KNAW), p.\ 5
\reference Papadopoulos, P. P., \etal 1999, \apj, in press
\reference Pentericci, L. \etal 1998, \aap, 341, 329
\reference Rengelink, R., \etal 1997, \aap, 124, 259
\reference R\"ottgering, H.J.A., Miley, G.K., Chambers, K.C. \& Macchetto, F. 1995, \aaps, 114, 51 
\reference R\"ottgering, H., van Ojik, R., Chambers, K., van Breugel, W., \& de Koff, S. 1997, \aap, 326, 505
\reference Shaver, P., Wall, J, Kellerman, K., Jackson, C., \& Hawkins, M. 1996, \nature, 384, 439
\reference Silk, J. \& Rees, M. 1998, \aap, 331, L1
\reference Spinrad, H., Filippenko, A. V., Wyckoff, S., Stocke, J. T., Wagner, R. M.,\& Lawrie, D. G. 1985, \apjl, 299, L7. 
\reference Spinrad, H., Dey, A. \& Graham, J. 1995, \apjl, 438, L51.
\reference Spinrad, H., Dey, A., Stern, D., Dunlop, J., Peacock, J., Jimenez, R. \& Windhorst, R. 1997, \apj, 484, 581 
\reference Spinrad, H. \etal 1998, \aj, 116, 2617
\reference Tielens, A. G. G. M., Miley, G. K., \& Willis, A. G. 1979, \aasup, 35, 153
\reference van Breugel, W., Stanford, S. A., Spinrad, H., Stern, D., \& Graham, J. R. 1998, \apj, 502, 614
\reference  van Breugel, W. \etal 1999, in 'The Most Distant Radio Galaxies', ed. H. R\"ottgering, P. Best \& M. Lehnert (Amsterdam: KNAW), p.\ 49
\reference van Ojik, R., R\"ottgering, H. J. A., Miley, G. K., \& Hunstead, R. W. 1997, \aap, 317, 358 
\reference Villar-Mart\'\i n, M., Tadhunter, C., \& Clark, N. 1997, \aap, 323, 21
\reference Weymann, R. \etal 1998, \apjl, 505, L95
\reference White, S. 1997, ESO-VLT Workshop "Galaxy Scaling Relations: Origins, Evolution and Applications", astro-ph/9702214
\end{references}
\end{document}